\documentclass[preprint,prd,aps,showpacs,showkeys,nofootinbib]{revtex4}
\usepackage{amssymb}
\usepackage{}
\usepackage{amsmath}
\usepackage{multirow}
\usepackage{graphicx}
\usepackage{float}

\begin{document}

\preprint{}

\title{The LFV decays of $Z$ boson in Minimal R-symmetric Supersymmetric Standard Model}

\author{Ke-Sheng Sun$^a$\footnote{sunkesheng@126.com;\;sunkesheng@mail.dlut.edu.cn}, Jian-Bin Chen$^{b}$\footnote{chenjianbin@tyut.edu.cn}, Xiu-Yi Yang$^{c}$\footnote{yxyruxi@163.com}, Sheng-Kai Cui$^d$\footnote{2252953633@qq.com}}

\affiliation{$^a$Department of Physics, Baoding University, Baoding 071000,China\\
$^b$College of Physics and Optoelectronic Engineering, Taiyuan University of Technology, Taiyuan 030024, China\\
$^c$School of Science, University of Science and Technology Liaoning, Anshan 114051, China\\
$^d$Department of Physics, Hebei University, Baoding 071002, China}

\begin{abstract}

A future $Z$-factory will offer the possibility to study rare $Z$ decays $Z\rightarrow l_1l_2$, as those
leading to Lepton Flavor Violation final states. In this work, by taking account of the constraints from radiative two body decays $l_2\rightarrow l_1\gamma$, we investigate the Lepton Flavor Violation decays $Z\rightarrow l_1l_2$ in the framework of Minimal R-symmetric Supersymmetric Standard Model with two benchmark points from already existing literatures. The flavor violating off-diagonal entries $\delta^{12}$, $\delta^{13}$ and $\delta^{23}$ are constrained by the current experimental bounds of $l_2\rightarrow l_1\gamma$. Considering recent experimental constraints, we also investigate Br($Z\rightarrow l_1l_2$) as a function of $M_D^W$. The numerical results show that the theoretical prediction of Br($Z\rightarrow l_1l_2$) in MRSSM are several orders of magnitude below the current experimental bounds. The Lepton Flavor Violation decays $Z\rightarrow e\tau$ and $Z\rightarrow \mu\tau$ may be promising to be observed in future experiment.
\end{abstract}



\maketitle

\section{Introduction}
\indent\indent

Rare decays are of great importance in searching for New Physics (NP) beyond the Standard Model (SM), and the Lepton Flavor Violating (LFV) decays are particularly appealing cause they are suppressed in SM, and their detection would be a manifest signal of NP. Search for such LFV decays has been pursued to date in a host of processes of leptons, $Z$ boson, Higgs boson and various hadrons. The present upper bounds on the various LFV decay channels of Z boson from both the LEP data and the LHC data is shown in TABLE.\ref{lfvZ} \cite{PDG}. One can find a pedagogical introduction on the theoretical motivations for charged LFV and the experimental aspects in Ref.\cite{Calibbi}.
\begin{table}[h]
\caption{Current limits on LFV decays of $Z$ boson. }
\begin{tabular}{@{}ccccc@{}} \toprule
Decay&Bound&Experiment&Bound&Experiment\\
\colrule
$Z\rightarrow e\mu$&$1.7\times 10^{-6}$&LEP (1995)\cite{ZPC9567}&$7.5\times 10^{-7}$&ATLAS (2014)\cite{AT14}\\
$Z\rightarrow e\tau$&$9.8\times 10^{-6}$&LEP (1995)\cite{ZPC9567}&$5.8\times 10^{-5}$&ATLAS (2018)\cite{AT18}\\
$Z\rightarrow \mu\tau$&$1.2\times 10^{-5}$&LEP (1995)\cite{ZPC9773}&$1.3\times 10^{-5}$&ATLAS (2018)\cite{AT18}\\
\botrule
\end{tabular}
\label{lfvZ}
\end{table}

As a new solution to the supersymmetric flavor problem in MSSM, the Minimal R-symmetric Supersymmetric Standard Model (MRSSM) is proposed in Ref.\cite{Kribs}, where the R-symmetry is a fundamental symmetry proposed several decades ago and stronger than R-parity \cite{Fayet,Salam}. R-symmetry forbids Majorana gaugino masses, $\mu$ term, $A$ terms and all left-right squark and slepton mass mixings. The $R$-charged Higgs $SU(2)_L$ doublets $\hat{R}_u$ and $\hat{R}_d$ are introduced in MRSSM to yield the Dirac mass terms of higgsinos.	Additional superfields $\hat{S}$, $\hat{T}$ and $\hat{O}$ are introduced to yield Dirac mass terms of gauginos. Studies on phenomenology in MRSSM can be found in literatures \cite{Die1, Die2, Die3, Die4, Die5, Kumar, Blechman, Kribs1, Frugiuele, Jan, Chakraborty, Braathen, Athron, Alvarado}.

In this paper, we have studied the LFV decays of $Z$ boson in MRSSM. Similar to the case in MSSM, the LFV decays mainly originate from the off-diagonal entries in slepton mass matrices $m_l^2$ and $m_r^2$ \cite{Kss}. Taking account of the constraint from radiative decay $l_2\rightarrow l_1\gamma$ on the off-diagonal parameters, we give the upper predictions on the LFV decays of $Z$ boson with parameter spaces BMP1 and BMP3\cite{Die3}. Taking account of recent experimental limit on the masses of charginos and neutrilinos \cite{AT1806}, we also explore the LFV decays of $Z$ boson as a function of Dirac mass parameter $M_D^W$. A comparison on the upper bounds of off-diagonal parameters between MRSSM and MSSM is also displayed.

The paper is organized as follows. In Section \ref{sec2}, we provide a brief introduction on MRSSM, and derive the analytic expressions for every Feynman diagram contributing to LFV decays of $Z$ boson in MRSSM in detail. The numerical results are presented in Section \ref{sec3}, and the conclusion is drawn in Section \ref{sec4}.

\section{Formalism\label{sec2}}

In this section, we firstly provide a simple overview of MRSSM. The spectrum of fields in MRSSM contains the standard MSSM matter, Higgs and gauge superfields augmented by chiral adjoints $\hat{\cal O},\hat{T},\hat{S}$ and two $R$-Higgs iso-doublets. The superfields with R-charge in MRSSM are given in TABLE.\ref{Field}.
\begin{table}[!htbp]
\centering
\caption{The superfields with R-charge in MRSSM. }
\begin{tabular}{|c|c|c|c|c|c|c|}
\hline
\multicolumn{1}{|c}{Field} & \multicolumn{2}{|c}{Superfield} &
                              \multicolumn{2}{|c}{Boson} &
                              \multicolumn{2}{|c|}{Fermion} \\
\hline
Gauge vector & $\hat{g},\hat{W},\hat{B}$&0& $g,W,B$&0& $\tilde{g},\tilde{W}\tilde{B}$&  +1 \\\hline
\multirow{2}*{Matter}& $\hat{l}, \hat{e}^c$& +1& $\tilde{l},\tilde{e}^*_R$&+1& $l,e^*_R$& 0   \\
 \cline{2-7}     & $\hat{q},{\hat{d}^c},{\hat{u}^c}$& +1& $\tilde{q},{\tilde{d}}^*_R,{\tilde{u}}^*_R$ & +1& $q,d^*_R,u^*_R$                             & 0 \\\hline
$H$-Higgs & ${\hat{H}}_{d,u}$&0& $H_{d,u}$& 0& ${\tilde{H}}_{d,u}$&-1 \\ \hline
R-Higgs & ${\hat{R}}_{d,u}$ & +2& $R_{d,u}$& +2& ${\tilde{R}}_{d,u}$& +1 \\\hline
Adjoint chiral& $\hat{\cal O},\hat{T},\hat{S}$&0& $O,T,S$&0& $\tilde{O},\tilde{T},\tilde{S}$ &-1 \\
\hline
\end{tabular}\label{Field}
\end{table}
The general form of the superpotential of the MRSSM is given by \cite{Die1}
\begin{eqnarray}
\mathcal{W}_{MRSSM} &=& \mu_d(\hat{R}_d\hat{H}_d)+\mu_u(\hat{R}_u\hat{H}_u)+\Lambda_d(\hat{R}_d\hat{T})\hat{H}_d+\Lambda_u(\hat{R}_u\hat{T})\hat{H}_u\nonumber\\
&+&\lambda_d\hat{S}(\hat{R}_d\hat{H}_d)+\lambda_u\hat{S}(\hat{R}_u\hat{H}_u)-Y_d\hat{d}(\hat{q}\hat{H}_d)-Y_e\hat{e}(\hat{l}\hat{H}_d)+Y_u\hat{u}(\hat{q}\hat{H}_u),
\end{eqnarray}
where $\hat{H}_u$ and $\hat{H}_d$ are the MSSM-like Higgs weak iso-doublets, $\hat{R}_u$ and $\hat{R}_d$ are the $R$-charged Higgs $SU(2)_L$ doublets and the corresponding Dirac higgsino mass parameters are denoted as $\mu_u$ and $\mu_d$. $\lambda_u$, $\lambda_d$, $\Lambda_u$ and $\Lambda_d$ are parameters of Yukawa-like trilinear terms involving the singlet $\hat{S}$ and the triplet $\hat{T}$, which is given by
\begin{equation}
\hat{T} = \left(
\begin{array}{cc}
\hat{T}^0/\sqrt{2} &\hat{T}^+ \\
\hat{T}^-  &-\hat{T}^0/\sqrt{2}\end{array}
\right).\nonumber
 \end{equation}
Then, the soft-breaking terms involving scalar mass are
\begin{eqnarray}
V_{SB,S} &=& m^2_{H_d}(|H^0_d|^2+|H^{-}_d|^2)+m^2_{H_u}(|H^0_u|^2+|H^{+}_u|^2)+(B_{\mu}(H^-_dH^+_u-H^0_dH^0_u)+h.c.)\nonumber\\
&+&m^2_{R_d}(|R^0_d|^2+|R^{+}_d|^2)+m^2_{R_u}(|R^0_u|^2+|R^{-}_u|^2)+m^2_T(|T^0|^2+|T^-|^2+|T^+|^2)\nonumber\\
&+&m^2_S|S|^2+ m^2_O|O^2|+\tilde{d}^*_{L,i} m_{q,{i j}}^{2} \tilde{d}_{L,j} +\tilde{d}^*_{R,i} m_{d,{i j}}^{2} \tilde{d}_{R,j}+\tilde{u}^*_{L,i}  m_{q,{i j}}^{2} \tilde{u}_{L,j}\nonumber\\
&+&\tilde{u}^*_{R,i}  m_{u,{i j}}^{2} \tilde{u}_{R,j}+\tilde{e}^*_{L,i} m_{l,{i j}}^{2} \tilde{e}_{L,j}+\tilde{e}^*_{R,{i}} m_{r,{i j}}^{2} \tilde{e}_{R,{j}} +\tilde{\nu}^*_{L,i} m_{l,{i j}}^{2} \tilde{\nu}_{L,j}.
\end{eqnarray}
It is noted that all trilinear scalar couplings involving Higgs bosons to squarks and sleptons are forbidden due to the $R$-symmetry. The Dirac nature is a manifest feature of MRSSM fermions and the soft-breaking Dirac mass terms of the singlet $\hat{S}$, triplet $\hat{T}$ and octet $\hat{O}$ take the form
\begin{equation}
V_{SB,DG}=M^B_D\tilde{B}\tilde{S}+M^W_D\tilde{W}^a\tilde{T}^a+M^O_D\tilde{g}\tilde{O}+h.c.,
\label{}
\end{equation}
where $\tilde{B}$, $\tilde{W}$ and $\tilde{g}$ are usually MSSM Weyl fermions. After EWSB, one can get the following $4\times 4$ neutralino mass matrix
\begin{eqnarray}
m_{\chi^0} &=& \left(
\begin{array}{cccc}
M^{B}_D &0 &-\frac{1}{2} g_1 v_d  &\frac{1}{2} g_1 v_u \\
0 &M^{W}_D &\frac{1}{2} g_2 v_d  &-\frac{1}{2} g_2 v_u \\
- \frac{1}{\sqrt{2}} \lambda_d v_d  &-\frac{1}{2} \Lambda_d v_d  &-\mu_d^{eff,+}&0\\
\frac{1}{\sqrt{2}} \lambda_u v_u  &-\frac{1}{2} \Lambda_u v_u  &0 &\mu_u^{eff,-}\end{array}
\right),
 \end{eqnarray}
where the modified $\mu_i$ parameters are
\begin{align}
\mu_d^{eff,+}&= \frac{1}{2} \Lambda_d v_T  + \frac{1}{\sqrt{2}} \lambda_d v_S  + \mu_d ,\nonumber\\
\mu_u^{eff,-}&= -\frac{1}{2} \Lambda_u v_T  + \frac{1}{\sqrt{2}} \lambda_u v_S  + \mu_u,\nonumber
\end{align}
and the $v_T$ and $v_S$ are vacuum expectation values of $\hat{T}$ and $\hat{S}$ which carry zero $R$-charge.
The neutralino mass matrix can be diagonalized by unitary matrices $N^1$ and $N^2$
\begin{equation}
(N^{1})^{\ast} m_{\chi^0} (N^{2})^{\dagger} = diag(m_{\chi^0_1},...,m_{\chi^0_4}).\nonumber
\end{equation}
The chargino mass matrix is given by
\begin{equation}
m_{\chi^{\pm}} = \left(
\begin{array}{cc}
g_2 v_T  + M^{W}_D &\frac{1}{\sqrt{2}} \Lambda_d v_d \\
\frac{1}{\sqrt{2}} g_2 v_d  &-\frac{1}{2} \Lambda_d v_T  + \frac{1}{\sqrt{2}} \lambda_d v_S  + \mu_d\end{array}
\right),
 \end{equation}
and can be diagonalized by unitary matrices $U^1$ and $V^1$
\begin{equation}
(U^{1})^{\ast} m_{\chi^{\pm}} (V^{1})^{\dagger} =diag(m_{\chi^{\pm}_1},m_{\chi^{\pm}_2}).\nonumber
\end{equation}
The LFV interactions mainly originate from the potential misalignment between the leptons and sleptons mass matrices in the MRSSM. In the gauge eigenstate basis $\tilde{\nu}_{iL}$, the sneutrino mass squared matrix is expressed as
\begin{equation}
m^2_{\tilde{\nu}} =
\begin{array}{c}m_l^2+\frac{1}{8}(g_1^2+g_2^2)( v_{d}^{2}- v_{u}^{2})+g_2 v_T M^{W}_D-g_1 v_S M^{B}_D,
\end{array}
 \end{equation}
where the last two terms are newly introduced by MRSSM, and the mass matrix is diagonalized by unitary matrix $Z^V$
\begin{equation}
Z^V m^2_{\tilde{\nu}} (Z^{V})^{\dagger} = diag(m^{2}_{\tilde{\nu}_1},m^{2}_{\tilde{\nu}_2},m^{2}_{\tilde{\nu}_3}).\nonumber
\end{equation}
The slepton mass squared matrix takes the form
\begin{equation}
m^2_{\tilde{L}^{\pm}} = \left(
\begin{array}{cc}
(m^2_{\tilde{L}^{\pm}})_{LL} &0 \\
0  &(m^2_{\tilde{L}^{\pm}})_{RR}\end{array}
\right),\label{sl}
 \end{equation}
where
\begin{align}
(m^2_{\tilde{L}^{\pm}})_{LL} &=m_l^2+ \frac{1}{2} v_{d}^{2} |Y_{e}|^2 +\frac{1}{8}(g_1^2-g_2^2)(v_{d}^{2}- v_{u}^{2}) -g_1 v_S M_D^B-g_2v_TM_D^W ,\nonumber\\
(m^2_{\tilde{L}^{\pm}})_{RR} &= m_r^2+\frac{v_d^2}{2}|Y_e|^2+\frac{1}{4}g_1^2( v_{u}^{2}- v_{d}^{2})+2g_1v_SM_D^B.\nonumber
\end{align}
The sources of LFV are the off-diagonal entries of the $3\times 3$ soft supersymmetry breaking matrices $m_l^2$ and $m_r^2$, where the A terms are absent. Note that, in the following, we replace $\tilde{l}$ with $\tilde{L}^{\pm}$ to denote the sleptons. From Eq.(\ref{sl}), we can see that the left-right slepton mass mixing is also absent. The slepton mass matrix is diagonalized by unitary matrix $Z^E$
\begin{equation}
Z^E m^2_{\tilde{L}^{\pm}} (Z^{E})^{\dagger} =diag(m^2_{\tilde{L}^{\pm}_1},...,m^2_{\tilde{L}^{\pm}_6}). \nonumber
\end{equation}

\begin{figure}[htbp]
\setlength{\unitlength}{1mm}
\centering
\begin{minipage}[c]{1\textwidth}
\includegraphics[width=5.0in]{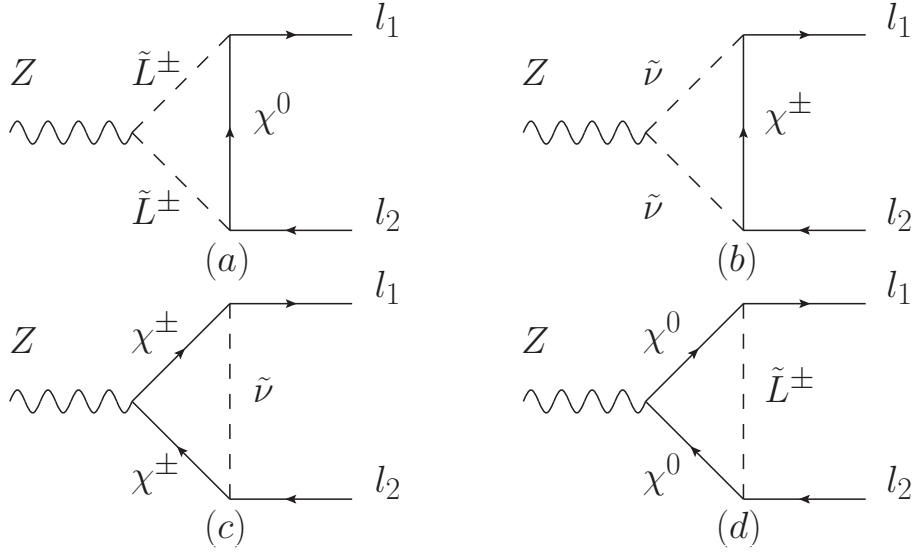}
\end{minipage}
\caption[]{One loop Feynman diagrams contributing to $Z\rightarrow \bar{l}_1l_2$ in MRSSM.}
\label{diag}
\end{figure}
The interactions of charged sleptons $\tilde{L}^{\pm}$ and neutral sneutrinos $\tilde{\nu}$ with neutralinos $\chi^0$ and charginos $\chi^{\pm}$ are correspondingly given by the Lagrangian as \cite{Die3,SARAH}
\begin{eqnarray}
-i\mathcal{L}&=&\bar{\chi}^0_i\Big[Y_e^{l_j}N^{2\ast}_{i3}Z^E_{k(3+j)}P_L+\sqrt{2}g_1Z^E_{k(3+j)}N^1_{i1}P_R\Big]l_j\tilde{L}^{\pm}_{k}\nonumber\\
&+&\bar{\chi}^{\pm}\Big[g_2V^{1\ast}_{i1}Z^V_{kj}P_L-Z^V_{kj}U^1_{i2}P_R\Big]l_j\tilde{\nu}_{k}.\nonumber
\end{eqnarray}
The interactions between $Z$ boson and neutralinos $\chi^0$ or charginos $\chi^{\pm}$ are given by the Lagrangian as
\begin{eqnarray}
-i\mathcal{L}&=&\frac{1}{2}\bar{\chi}^{\pm}_{i}\gamma_{\mu}\Big[\Big(2g_2c_wV^{1\ast}_{j1}V^1_{i1}+(g_2c_w-g_1s_w)V^{1\ast}_{j2}V^1_{i2}\Big)P_L
+\Big((g_2c_w-g_1s_w)U^{1\ast}_{i2}U^1_{j2}\nonumber\\
&+&2g_2c_wU^{1\ast}_{i1}U^1_{j2}\Big)P_R\Big]\tilde{\chi}^{\pm}_{j}Z^{\mu}\nonumber\\
&+&\frac{1}{2}(g_1s_w+g_2 c_w)\bar{\chi}^0_i\gamma_{\mu}\Big[(N^{1\ast}_{j4}N^{1}_{i4}-N^{1\ast}_{j3}N^1_{i3})P_L
+(N^{2\ast}_{i4}N^{2}_{j4}-N^{2\ast}_{i3}N^2_{j3})P_R\Big]\chi^0_jZ^{\mu}.
\nonumber
\end{eqnarray}

The relevant Feynman diagrams contributing to the LFV decays of $Z$ boson in MRSSM is presented in FIG.\ref{diag}, where FIG.\ref{diag}(a) and FIG.\ref{diag}(c) take more important role than others. The $Z l_1l_2$ interaction Lagrangian can be written as \cite{Flavor}
\begin{eqnarray}
\mathcal{L}_{Z l_1l_2}=\bar{l}_1\Big[\gamma^{\mu}(C^1_L P_L+ C^1_R P_R)+p_1^{\mu}(C^2_L P_L+ C^2_R P_R)\Big]l_2Z_{\mu}.
\end{eqnarray}
The left-handed current coefficient $C^1_L$ for FIG.\ref{diag}(a) and the right-handed current coefficient $C^1_R$ for FIG.\ref{diag}(c) are dominant in the final result. Then the branching ratios of LFV decays of $Z$ boson is calculated by
\begin{eqnarray}
Br(Z\rightarrow l_1 l_2)&=&Br(Z\rightarrow \bar{l}_1 l_2)+Br(Z\rightarrow \bar{l}_2 l_1)\nonumber\\
&=&\frac{m_Z}{48\pi\Gamma_Z}\Big[2(|C^1_L|^2+|C^1_R|^2)+\frac{m_Z^2}{4}(|C^2_L|^2+|C^2_R|^2)\Big],
\end{eqnarray}
where the charged lepton masses have been neglected and $\Gamma_Z$ is the total decay width of $Z$ boson. The coefficients $C^{1}_{L/R}$ and $C^{2}_{L/R}$ are combinations of coefficients corresponding to each Feynman diagram in FIG.\ref{diag} and take the form
\begin{eqnarray}
C^{1}_{L/R}&=&C^{1,a}_{L/R}+C^{1,b}_{L/R}+C^{1,c}_{L/R}+C^{1,d}_{L/R},\nonumber\\
C^{2}_{L/R}&=&C^{2,a}_{L/R}+C^{2,b}_{L/R}+C^{2,c}_{L/R}+C^{2,d}_{L/R}.\nonumber
\end{eqnarray}
The coefficients in FIG.\ref{diag} (a) and FIG.\ref{diag} (b) take the same form
\begin{eqnarray}
C_L^{1,a/b}&=&2C_{1R}C_2C_{3L}\mathcal{C}_{00},C_R^{1,a/b}=2C_{1L}C_2C_{3R}\mathcal{C}_{00},\nonumber\\
C_L^{2,a/b}&=&2 C_2\Big(C_{1L}(C_{3R}m_2-C_{3L}M_3)\mathcal{C}_{2}-C_{1L}C_{3L}M_3\mathcal{C}_0-C_{1L}C_{3L}M_3\mathcal{C}_1\nonumber\\
&+&C_{1L}C_{3R}m_2\mathcal{C}_{22}+C_{1L}C_{3R}m_2\mathcal{C}_{12}+C_{1R}C_{3L}m_1\mathcal{C}_{12}+C_{1R}C_{3L}m_1\mathcal{C}_{11}\nonumber\\
&+&C_{1R}C_{3L}m_1\mathcal{C}_1\Big),C_R^{2,a/b}=C_L^{1,a/b}(L\leftrightarrow R),\nonumber
\end{eqnarray}
where the couplings corresponding to FIG.\ref{diag} (a) is
\begin{eqnarray}
C^{(a)}_{1L}&=&-i\sqrt{2}g_1N^{1\ast}_{j1}Z^{E\ast}_{k(3+l_1)},C^{(a)}_{1R}=-iY_e^{l_1}Z^{E\ast}_{k(3+l_1)}N^2_{j3},\nonumber\\
C^{(a)}_{2}&=&i/2\Big((g_2 c_w-g_1 s_w)\sum_{l=1,2,3}Z^{E\ast}_{il}Z^{E}_{kl}-2g_1 s_w \sum_{l=1,2,3}Z^{E\ast}_{i(3+l)}Z^{E}_{k(3+l)}\Big),\nonumber\\
C^{(a)}_{3L}&=&-iY_e^{l_2}N^{2\ast}_{j3}Z^E_{i(3+l_2)},C^{(a)}_{3R}=-i\sqrt{2}g_1Z^E_{i(3+l_2)}N^1_{j1},\nonumber\\
M_1&=&m_{\tilde{L}^{\pm}_k},M_2=m_{\tilde{L}^{\pm}_i},M_3=m_{\chi^{0}_j},\nonumber
\end{eqnarray}
and the couplings corresponding to FIG.\ref{diag} (b) is
\begin{eqnarray}
C^{(b)}_{1L}&=&iY_e^{l_1}U^{1\ast}_{j2}Z^{V\ast}_{kl_1},C^{(b)}_{1R}=-ig_2Z^{V\ast}_{kl_1}V^1_{j1},
C^{(b)}_{2}=-i/2(g_2 c_w+g_1 s_w),\nonumber\\
C^{(b)}_{3L}&=&-ig_2Z^{V}_{il_2}V^{1\ast}_{j1},C^{(b)}_{3R}=iY_e^{l_2}U^1_{j2}Z^V_{il_2},
M_1=m_{\tilde{\nu}_k},M_2=m_{\tilde{\nu}_i},M_3=m_{\chi^{\pm}_j}.\nonumber
\end{eqnarray}
The coefficients in FIG.\ref{diag} (c) and FIG.\ref{diag} (d)  take the same form
\begin{eqnarray}
C_L^{1,c/d}&=&C_{1R}C_{2L}C_{3L}\mathcal{B}_0-m_2\mathcal{C}_0(C_{1L}C_{2L}C_{3R}m_1+C_{1R}(C_{2L}C_{3R}M_1-C_{2R}C_{3L}m_2\nonumber\\
&+&C_{2R}C_{3R}M_2))-C_{1R}C_{2L}C_{3R}M_1m_2\mathcal{C}_0+C_{1R}C_{2R}C_{3L}M_3^2\mathcal{C}_0-2C_{1R}C_{2R}C_{3L}\mathcal{C}_{00}\nonumber\\
&+&C_{1L}C_{2L}C_{3L}m_1M_2\mathcal{C}_0-C_{1L}C_{2L}C_{3R}m_1m_2\mathcal{C}_0+C_{1R}C_{2L}C_{3L}M_1M_2\mathcal{C}_0\nonumber\\
&+&m_1\mathcal{C}_1(C_{1L}(C_{2L}C_{3L}M_2-C_{2L}C_{3R}m_2+C_{2R}C_{3L}M_1)+C_{1R}C_{2R}C_{3L}m_1),\nonumber\\
C_R^{1,c/d}&=&C_L^{1,c/d}(L\leftrightarrow R),C_R^{2,c/d}=C_L^{2,c/d}(L\leftrightarrow R),\nonumber\\
C_L^{2,c/d}&=&2(C_{1L}C_{2L}(C_{3L}M_2-C_{3R}m_2)\mathcal{C}_2-C_{1L}C_{2L}C_{3R}m_2\mathcal{C}_{22}-C_{1L}C_{2L}C_{3R}m_2\mathcal{C}_{12}\nonumber\\
&-&(C_{1L}M_1+C_{1R}m_1)C_{2R}C_{3L}\mathcal{C}_1-C_{1R}C_{2R}C_{3L}m_1\mathcal{C}_{12}-C_{1R}C_{2R}C_{3L}m_1\mathcal{C}_{11}),\nonumber
\end{eqnarray}
where the couplings corresponding to FIG.\ref{diag} (c) is
\begin{eqnarray}
C^{(c)}_{1L}&=&iY_e^{l_1}U^{1\ast}_{k2}Z^{V\ast}_{jl_1},C^{(c)}_{1R}=-ig_2Z^{V\ast}_{jl_1}V^1_{k1},\nonumber\\
C^{(c)}_{2L}&=&-i/2((g_2 c_w-g_1 s_w)V^{1\ast}_{i2}V^1_{k2}+2g_2 c_wV^{1\ast}_{i1}V^1_{k1}),C^{(c)}_{3L}
=-ig_2Z^V_{jl_2}V^{1\ast}_{i1},\nonumber\\
C^{(c)}_{2R}&=&-i/2((g_2 c_w-g_1 s_w)U^{1\ast}_{k2}U^1_{i2}+2g_2 c_wU^{1\ast}_{k1}U^1_{i1}),C^{(c)}_{3R}=iY_e^{l_2}U^1_{i2}Z^V_{jl_2},\nonumber\\
M_1&=&m_{\chi^{\pm}_k},M_2=m_{\chi^{\pm}_i},M_3=m_{\tilde{\nu}_j}.\nonumber
\end{eqnarray}
and the couplings corresponding to FIG.\ref{diag} (d) is
\begin{eqnarray}
C^{(d)}_{1L}&=&-i\sqrt{2}g_1N^{1\ast}_{k1}Z^{E\ast}_{j(3+l_1)},C^{(d)}_{1R}=-iY_e^{l_1}Z^{E\ast}_{j(3+l_1)}N^2_{k3},\nonumber\\
C^{(d)}_{2L}&=&i/2(g_2 c_w+g_1 s_w)(N^{1\ast}_{i3}N^{1}_{k3}-N^{1\ast}_{i4}N^{1}_{k4}),C^{(d)}_{3L}=-iY_e^{l_2}N^{2\ast}_{i3}Z^E_{j(3+l_2)},\nonumber\\
C^{(d)}_{2L}&=&i/2(g_2 c_w+g_1 s_w)(N^{2\ast}_{k3}N^{2}_{i3}-N^{2\ast}_{k4}N^{2}_{i4}),C^{(d)}_{3R}=-i\sqrt{2}g_1Z^E_{j(3+l_2)}N^1_{i1},\nonumber\\
M_1&=&m_{\chi^{0}_k},M_2=m_{\chi^{0}_i},M_3=m_{\tilde{L}^{\pm}_j}.\nonumber
\end{eqnarray}
Above loop integrals are given in term of Passarino-Veltman functions \cite{PVI}
\begin{eqnarray}
\mathcal{B}_0&=&\frac{i}{16\pi^2}\mathcal{B}_{0}(m_Z^2,M_1,M_2),\nonumber\\
\mathcal{C}_{1,2,...}&=&\frac{i}{16\pi^2}\mathcal{C}_{1,2,...}(m_1^2,m_Z^2,m_2^2;M_3,M_1,M_2),\nonumber
\end{eqnarray}
and can be calculated by the Mathematica package Package-X \cite{X} through a link to fortran library Collier \cite{collier,collier1,collier2,collier3}, where the latter provides the numerical evaluation of one-loop scalar and tensor integrals in perturbative relativistic quantum field theories.

\section{Numerical Analysis\label{sec3}}
\indent\indent
In the numerical analysis, we use the benchmark points in Ref.\cite{Die3} as the default values for our parameter setup and display them in Table.\ref{BMP}, where the slepton mass matrices are diagonal and all mass parameters are in $GeV$ or $GeV^2$ and the mass spectra for the BMPs are shown in Table.\ref{MASS}. Note that large value of $|v_T|$ is excluded by measurement of $W$ mass cause the vev $v_T$ of the $SU(2)_L$ triplet field $T^0$ gives a correction to $W$ mass through \cite{Die1}
\begin{eqnarray}
m_W^2=\frac{1}{4}g_2^2(v_u^2+v_d^2)+g_2^2v_T^2.
\label{}
\end{eqnarray}

\begin{table}[h]
\caption{Benchmark points. }
\begin{tabular}{@{}ccccccccccc@{}}
\toprule
Input&$tan\beta$&$\lambda_d$,$\lambda_u$&$\Lambda_d$,$\Lambda_u$&$v_S$&$v_T$&$M_D^B$&$M_D^W$&$\mu_d$,$\mu_u$&$m_T^2$&$m^2_l$,$m_r^2$\\
\colrule
BMP1&3&1.0,-0.8&-1.0,-1.2&5.9&-0.33&600&500&400,400&$3000^2$&$1000^2$,$1000^2$\\
\colrule
BMP3&40&0.15,-0.15&-1.0,-1.15&-0.14&-0.34&250&500&400,400&$3000^2$&$1000^2$,$1000^2$\\
 \botrule
\end{tabular}\label{BMP}
\end{table}

\begin{table}[h]
\caption{Spectrum of the different BMPs in GeV. }
\begin{tabular}{@{}ccccccccc@{}}
\toprule
Input&$H_1$&$H_2$&$A_1$&$H^{\pm}_1$&$\chi^0_1$&$\chi^0_2$&$\chi^{\pm}_1$&$\tilde{\nu}$\\
\colrule
BMP1&125.3&897&896&899&415&420&416&1002\\
\colrule
BMP3&125.1&1245&1245&1248&251&408&408&1000\\
 \botrule
\end{tabular}\label{MASS}
\end{table}

To decrease the number of free parameters involved in our calculation,
we assume that the diagonal entries of two $3\times3$ matrices $m_{l}^{2}$ and $m_{r}^{2}$
are equal $(m_{l}^{2})_{II}=(m_{r}^{2})_{II}$, same to the values shown in TABLE.\ref{BMP}, where
$I=1,2, 3$. Then, the only sources of LFV are off-diagonal entries of the soft breaking terms
$m_{l}^{2}$, $m_{r}^{2}$, which are parameterized by mass insertion as in \cite{Rosiek3}
\begin{eqnarray}
\Big(m^{2}_{l}\Big)^{IJ}&=&\delta ^{IJ}_{l}\sqrt{(m^{2}_{l})^{II}(m^{2}_{l})^{JJ}},\nonumber\\
\Big(m^{2}_{r}\Big)^{IJ}&=&\delta ^{IJ}_{r}\sqrt{(m^{2}_{r})^{II}(m^{2}_{r})^{JJ}},\nonumber
\end{eqnarray}
where $I,J=1,2,3$. We also assume $\delta ^{IJ}_{l}$ = $\delta ^{IJ}_{r}$ = $\delta ^{IJ}$. The experimental limits on LFV decays, such as radiative two body decays $l_2\rightarrow l_1\gamma$, leptonic three body decays $l_2\rightarrow 3l_1$ and $\mu-e$ conversion in nuclei, can give strong constraints on the parameters $\delta ^{IJ}$. In the following, we will use LFV decays $l_2\rightarrow l_1\gamma$ to constrain the parameters $\delta ^{IJ}$. Current limits of LFV decays $l_2\rightarrow l_1\gamma$ listed in TABLE.\ref{lfvl} \cite{PDG}.
\begin{table}[h]
\caption{Current limits of LFV decays of $l_2\rightarrow l_1\gamma$. }
\begin{tabular}{@{}cccccc@{}} \toprule
Decay&Bound&Experiment&Decay&Bound&Experiment\\
\colrule
$\mu\rightarrow e\gamma$&$4.2\times 10^{-13}$&MEG(2016)\cite{MEG}&$\tau\rightarrow e\gamma$&$3.3\times 10^{-8}$&BABAR(2010)\cite{BABAR}\\
$\tau\rightarrow \mu\gamma$&$4.4\times 10^{-8}$&BABAR(2010)\cite{BABAR}&-&-&-\\ \botrule
\end{tabular}
\label{lfvl}
\end{table}

\begin{figure}[htbp]
\setlength{\unitlength}{1mm}
\centering
\begin{minipage}[c]{1\textwidth}
\includegraphics[width=5.0in]{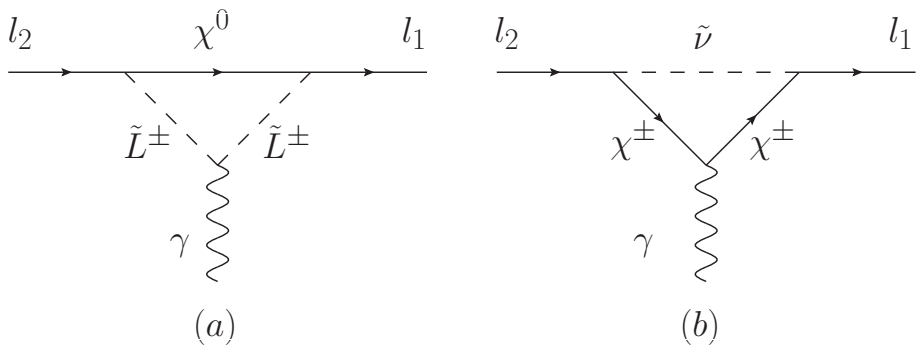}
\end{minipage}
\caption[]{Feynman diagrams contributing to the LFV decays $l_2\rightarrow l_1\gamma$ in MRSSM.}
\label{lfv}
\end{figure}

The sparticle mediated diagrams for $l_2\rightarrow l_1\gamma$ in MRSSM are shown in FIG.\ref{lfv}. Taking account of the gauge invariance, and assuming the photon is on shell and transverse, the amplitude for $l_2\rightarrow l_1\gamma$ is given by \cite{DingY}
\begin{eqnarray}
M(l_2\rightarrow l_1\gamma)=\epsilon^{\mu\ast}\bar{u}_{l_1}(p_{l_1})[iq^{\nu}\sigma_{\mu\nu}(A+B\gamma_5)]u_{l_2}(p_{l_2}).
\end{eqnarray}
Then, in the limit $m_1\rightarrow 0$, the analytic expression of $Br(l_2\rightarrow l_1\gamma)$ is derived as
\begin{eqnarray}
Br(l_2\rightarrow l_1\gamma)=\frac{m_{l_2}^3}{8\pi \Gamma_{l_2}}(|A|^2+|B|^2),
\end{eqnarray}
where $\Gamma_{l_2}$ is the total decay width of $l_2$ and the form factors $A$ and $B$ is a combination of form factors for every Feynman diagram in FIG.\ref{lfv},
\begin{eqnarray}
A=A^{(a)}+A^{(b)},B=B^{(a)}+B^{(b)}.\nonumber
\label{muABb}
\end{eqnarray}
In the limit $m_1\rightarrow 0$, the form factors $A^{(a)}$ and $B^{(a)}$ corresponding to FIG.\ref{lfv} (a) are given as
\begin{eqnarray}
A^{(a)}&=&-\frac{i}{32\pi^2}\Big[m_2(C_{1L}C_{2L}C_{3R}+C_{1R}C_{2R}C_{3L})(\mathcal{C}_{12}+\mathcal{C}_{11})+M_1(C_{1L}C_{2R}C_{3L}\nonumber\\
&+&C_{1R}C_{2L}C_{3R})\mathcal{C}_{2}+(M_2(C_{1L}C_{2L}C_{3L}+C_{1R}C_{2R}C_{3R})+m_2(C_{1L}C_{2L}C_{3R}\nonumber\\
&+&C_{1R}C_{2R}C_{3L}))\mathcal{C}_1\Big],\nonumber\\
B^{(a)}&=&\frac{i}{32\pi^2}\Big[M_1(C_{1L}C_{2R}C_{3L}-C_{1R}C_{2L}C_{3R})\mathcal{C}_2-m_2(C_{1R}C_{2R}C_{3L}-C_{1L}C_{2L}C_{3R})\nonumber\\
&\times&(\mathcal{C}_{12}+\mathcal{C}_{11})+(C_{1L}C_{2L}(M_2C_{3L}+m_2C_{3R})-C_{1R}C_{2R}(M_2C_{3R}+m_2C_{3L}))\mathcal{C}_1\Big],\nonumber
\label{muABa}
\end{eqnarray}
and the form factors $A^{(b)}$ and $B^{(b)}$ corresponding to FIG.\ref{lfv} (b) are given as
\begin{eqnarray}
A^{(b)}&=&-\frac{iC_2}{64\pi^2}\Big[M_3(C_{1L}C_{3L}+C_{1R}C_{3R})(\mathcal{C}_0+2\mathcal{C}_1+2\mathcal{C}_2)-m_2(C_{1L}C_{3R}+C_{1R}C_{3L})\nonumber\\
&\times&(2\mathcal{C}_{12}+2\mathcal{C}_{11}+\mathcal{C}_{1})\Big],\nonumber\\
B^{(b)}&=&\frac{iC_2}{64\pi^2}\Big[M_3(C_{1L}C_{3L}-C_{1R}C_{3R})(\mathcal{C}_0+2\mathcal{C}_1+2\mathcal{C}_2)-m_2(C_{1L}C_{3R}-C_{1R}C_{3L})\nonumber\\
&\times&(2\mathcal{C}_{12}+2\mathcal{C}_{11}+\mathcal{C}_{1})\Big],\nonumber
\label{muABb}
\end{eqnarray}
where the loop integrals $\mathcal{C}_{1,2,...}$ denote $\frac{i}{16\pi^2}\mathcal{C}_{1,2,...}(m_2^2,0,0;M_3,M_2,M_1)$. The couplings corresponding to FIG.\ref{lfv} (a) are given as
\begin{eqnarray}
C^{(a)}_{1L}&=&iY_e^{l_1}U^{1\ast}_{i2}Z^V_{kl_1},C^{(a)}_{1R}=-ig_2Z^{V\ast}_{kl_1}V^1_{i1},C^{(a)}_{2L}=C^{(a)}_{2R}=-ie\delta_{ij},\nonumber\\
C^{(a)}_{3L}&=&-ig_2V^{1\ast}_{j1}Z^V_{kl_2},
C^{(a)}_{3R}=iY_e^{l_2}U^{1}_{j2}Z^V_{kl_2},M_1=m_{\chi^{\pm}_i},M_2=m_{\chi^{\pm}_j},M_3=m_{\tilde{\nu}_k}.\nonumber
\label{ABCa}
\end{eqnarray}
and the couplings corresponding to FIG.\ref{lfv} (b) are given as
\begin{eqnarray}
C^{(b)}_{1L}&=&-i\sqrt{2}g_1N^{1\ast}_{k1}Z^{E\ast}_{i(3+l_1)},C^{(b)}_{1R}=-iY_e^{l_1}Z^{E\ast}_{i(3+l_1)}N^{2}_{k3},
C^{(b)}_{3L}=-iY_e^{l_2}N^{2\ast}_{k3}Z^E_{j(3+l_2)},\nonumber\\
C^{(b)}_2&=&ie\delta_{ij},C^{(b)}_{3R}=-i\sqrt{2}g_1Z^E_{j(3+l_2)}N^1_{k1},
M_1=m_{\tilde{L}^{\pm}_{i}},M_2=m_{\tilde{L}^{\pm}_{j}},M_3=m_{\chi^{0}_k}.\nonumber
\label{ABCb}
\end{eqnarray}

\begin{figure}[htbp]
\setlength{\unitlength}{1mm}
\centering
\begin{minipage}[c]{0.95\columnwidth}
\includegraphics[width=1\columnwidth]{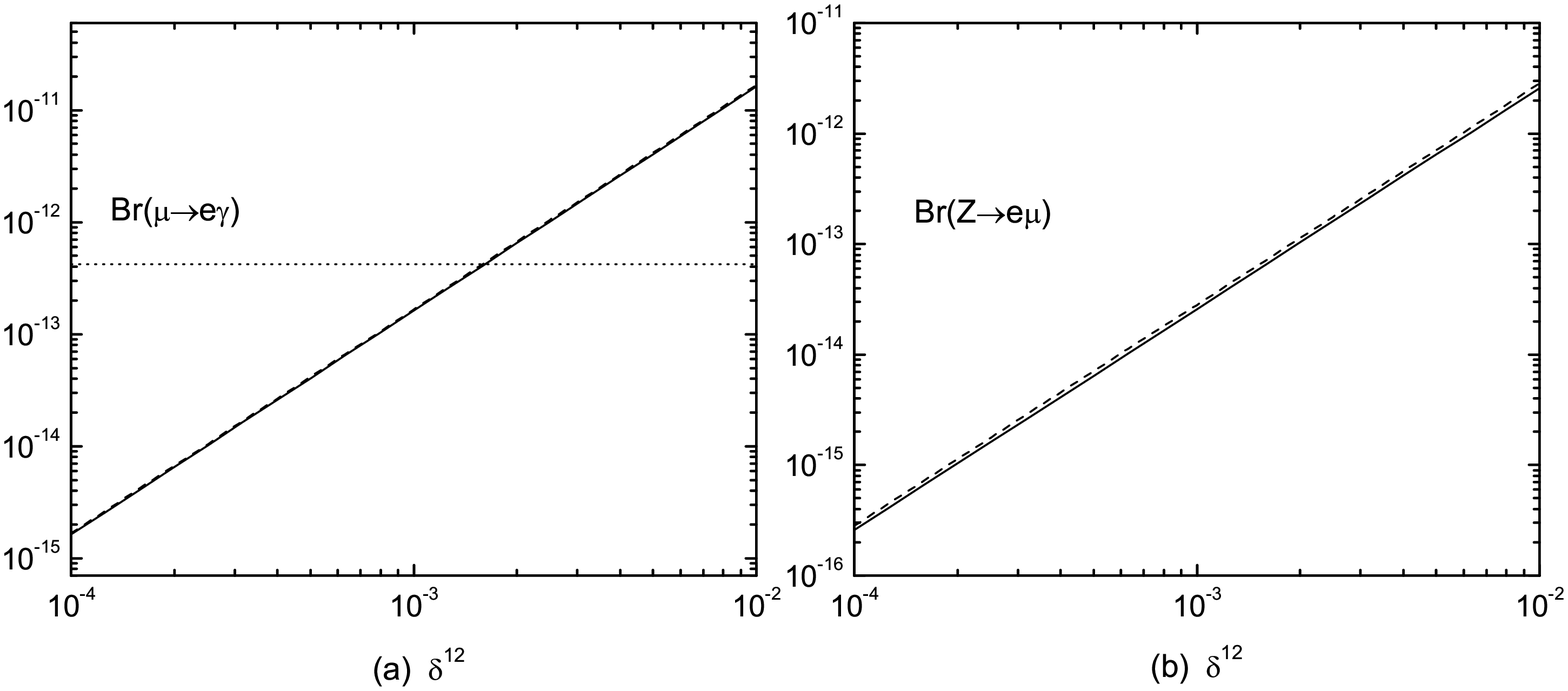}
\includegraphics[width=1\columnwidth]{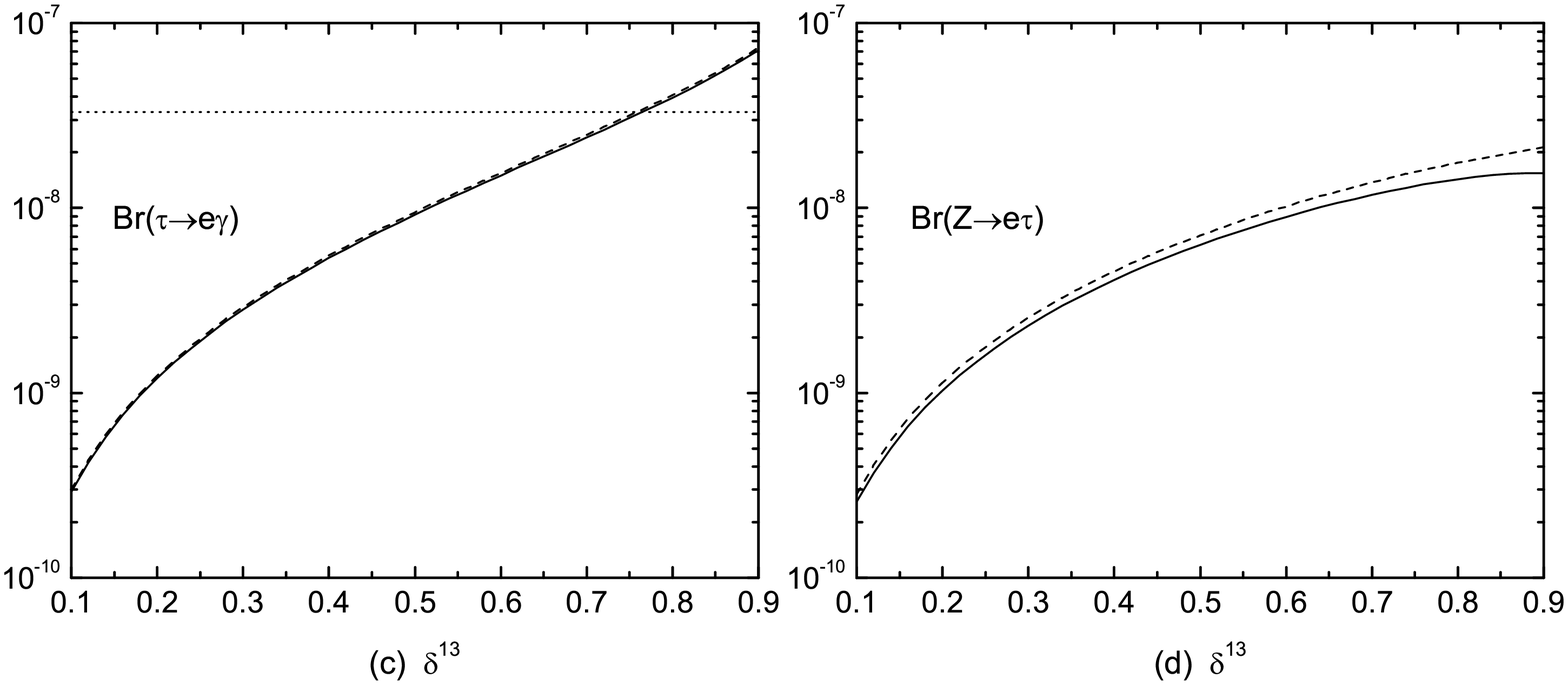}
\includegraphics[width=1\columnwidth]{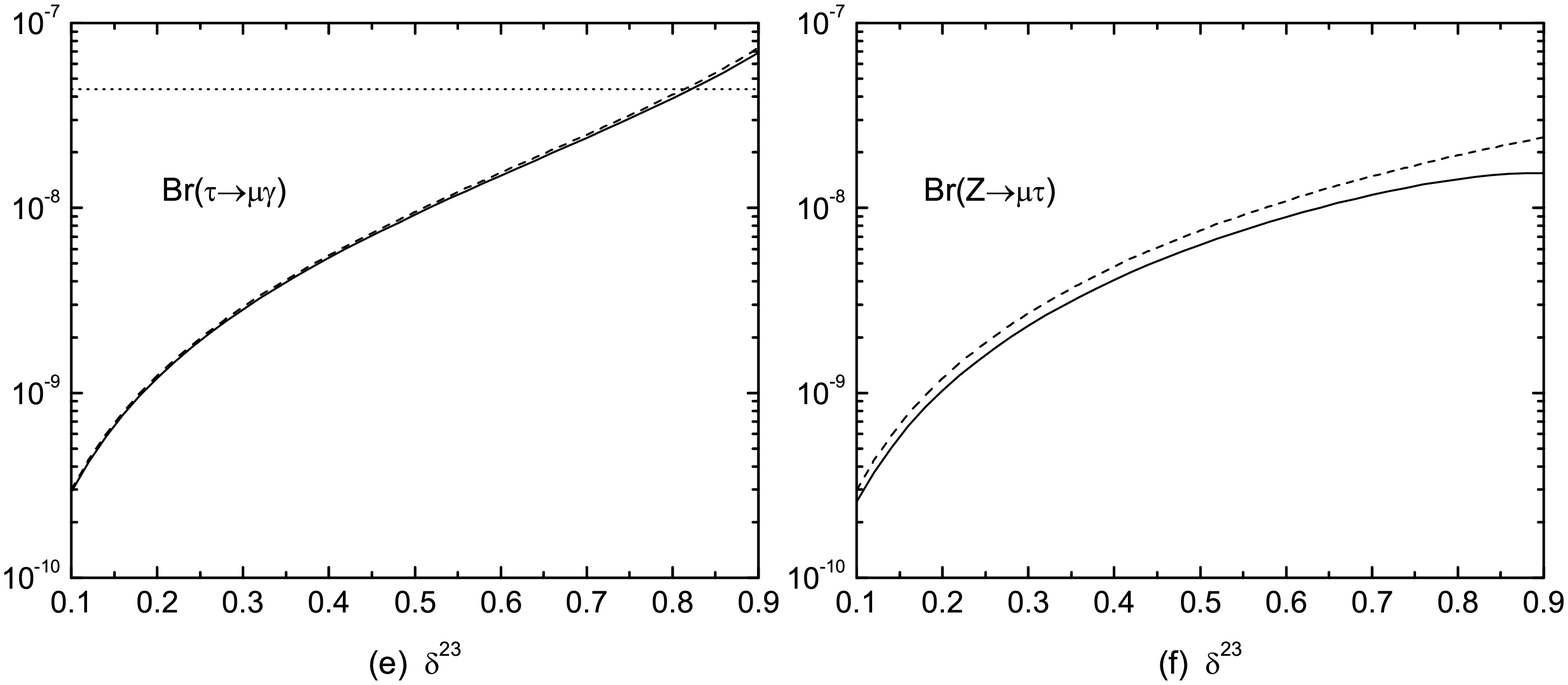}
\end{minipage}
\caption[]{(a)Br($\mu\rightarrow e\gamma$) versus $\delta^{12}$; (b)Br($Z\rightarrow e\mu$) versus $\delta^{12}$; (c)Br($\tau\rightarrow e\gamma$) versus $\delta^{13}$; (d)Br($Z\rightarrow e\tau$) versus $\delta^{13}$; (e)Br($\tau\rightarrow \mu\gamma$) versus $\delta^{23}$; (f)Br($Z\rightarrow \mu\tau$) versus $\delta^{23}$. The horizontal dot line in left panel denotes the current experimental bounds. The solid line stands for the result with BMP1 and the dash line stands for the result with BMP3.}
\label{figs1}
\end{figure}
Taking $\delta^{13}=0$, $\delta^{23}=0$, we plot the theoretical prediction of Br $(\mu\rightarrow e\gamma)$ versus $\delta^{12}$ and Br($Z\rightarrow e\mu$) versus $\delta^{12}$ in FIG.\ref{figs1}(a) and FIG.\ref{figs1}(b), where the horizontal dot line is the current experimental bounds of Br($\mu\rightarrow e\gamma$). The solid line stands for the result calculated with parameter setup BMP1 and the dash line stands for the result calculated with parameter setup BMP3. A linear relationship in logarithmic scale is displayed between Br$(\mu\rightarrow e\gamma)$ or Br($Z\rightarrow e\mu$) and the flavor violating parameter $\delta^{12}$. The prediction on Br$(\mu\rightarrow e\gamma)$ exceeds the current experiment limit at $\delta^{12}\sim 1.0\times 10^{-3}$. The parameter space of $\delta^{12}$ would been highly suppressed below to $\mathcal{O}(10^{-3})$ by taking account of the future sensitivity of experiment, which is Br$(\mu\rightarrow e\gamma)\sim 6 \times 10^{-14}$ \cite{MEG1}. The current prediction of Br($Z\rightarrow e\mu$) in MRSSM is around $\mathcal{O}(10^{-13})$ and this prediction is six orders of magnitude below the current limit $\mathcal{O}(10^{-7})$. Based on the flavour expansion theorem, various LFV processes have been investigated in MSSM using a recently developed technique which performes a purely algebraic mass-insertion expansion of the amplitudes \cite{Rosiek1}. Taking account of the constraints from radiative charged lepton decays, upper bounds on the flavor violating parameters $\delta^{12}_l $ ($\Delta^{21}_{LL} $) and $\delta^{12}_r $ ($\Delta^{21}_{RR} $) are given in Ref.\cite{Rosiek1} with $\delta^{12}_l \sim 8.4\times 10^{-4}$ ($\tan \beta$=2) and $\delta^{12}_r \sim 5.0\times 10^{-3}$ ($\tan \beta$=2), and it shows flavor violating $\mu$ lepton decays still provide the most stringent bounds on supersymmetric effects.

Taking $\delta^{12}=0$, $\delta^{23}=0$, we plot the theoretical prediction of Br $(\tau\rightarrow e\gamma)$ versus $\delta^{13}$ and Br($Z\rightarrow e\tau$) versus $\delta^{13}$ in FIG.\ref{figs1}(d) and FIG.\ref{figs1}(d), where the horizontal dot line is the current experimental bounds of Br($\tau\rightarrow e\gamma$). The solid line stands for the result calculated with parameter setup BMP1 and the dash line stands for the result calculated with parameter setup BMP3. Both predictions on Br$(\tau\rightarrow e\gamma)$ and Br($Z\rightarrow e\tau$) decrease as the flavor violating parameter $\delta^{13}$ varies from 0.9 to 0.1. The prediction on Br$(\tau\rightarrow e\gamma)$ exceeds the current experiment limit at $\delta^{13}\sim 0.75$. Taking account of the future experimental expectation on Br$(\tau\rightarrow e\gamma)$, which is around $2.3\times 10^{-9}$ \cite{SuperB}, the parameter space of $\delta^{13}$ would been suppressed below to $0.3$. The upper theoretical prediction on Br($Z\rightarrow e\tau$) in MRSSM is around $\mathcal{O}(10^{-8})$ and this prediction is two orders of magnitude below the current limit $\mathcal{O}(10^{-6})$. Recent upper bounds on the flavor violating parameters $\delta^{13}_l $ ($\Delta^{31}_{LL} $) and $\delta^{13}_r $ ($\Delta^{31}_{RR} $) in MSSM are given by $\delta^{13}_l \sim 4.6\times 10^{-1}$ ($\tan \beta$=2) and $\delta^{13}_r \sim \mathcal{O}(1)$ ($\tan \beta$=2) \cite{Rosiek1}.

Taking $\delta^{12}=0$, $\delta^{13}=0$, we plot the theoretical prediction of Br $(\tau\rightarrow \mu\gamma)$ versus $\delta^{23}$ and Br($Z\rightarrow \mu\tau$) versus $\delta^{23}$ in FIG.\ref{figs1}(e) and FIG.\ref{figs1}(f), where the horizontal dot line is the current experimental bounds of Br($\tau\rightarrow \mu\gamma$). The solid line stands for the result calculated with parameter setup BMP1 and the dash line stands for the result calculated with parameter setup BMP3. Both predictions on Br$(\tau\rightarrow \mu\gamma)$ and Br($Z\rightarrow \mu\tau$) decrease sharply as the flavor violating parameter $\delta^{13}$ is close to 0.1. The prediction on Br$(\tau\rightarrow \mu\gamma)$ exceeds the current experiment limit at $\delta^{23}\sim 0.8$. Taking account of the future experimental expectation on Br$(\tau\rightarrow \mu\gamma)$, which is around ($1.8\times 10^{-9}$) \cite{SuperB}, the parameter space of $\delta^{23}$ would been also suppressed below to $0.3$. The upper theoretical prediction on Br($Z\rightarrow \mu\tau$) in MRSSM is also around $\mathcal{O}(10^{-8})$ and this prediction is three orders of magnitude below the current limit $\mathcal{O}(10^{-5})$. Recent upper bounds on the flavor violating parameters $\delta^{23}_l $ ($\Delta^{32}_{LL} $) and $\delta^{23}_r $ ($\Delta^{32}_{RR} $) in MSSM are given by $\delta^{23}_l \sim 5.3\times 10^{-1}$ ($\tan \beta$=2) and $\delta^{23}_r \sim \mathcal{O}(1)$ ($\tan \beta$=2) \cite{Rosiek1}.

\begin{table}[h]
\caption{The range of input parameters for the numerical scan with BMP3. }
\begin{tabular}{@{}cccc@{}} \toprule
Parameter&Min&Max&Step\\
\colrule
$M_D^W$&600&800&10\\
$M_D^B$&600&1000&10\\
\botrule
\end{tabular}
\label{scan}
\end{table}
\begin{figure}[htbp]
\setlength{\unitlength}{1mm}
\centering
\begin{minipage}[c]{0.95\columnwidth}
\includegraphics[width=0.5\columnwidth]{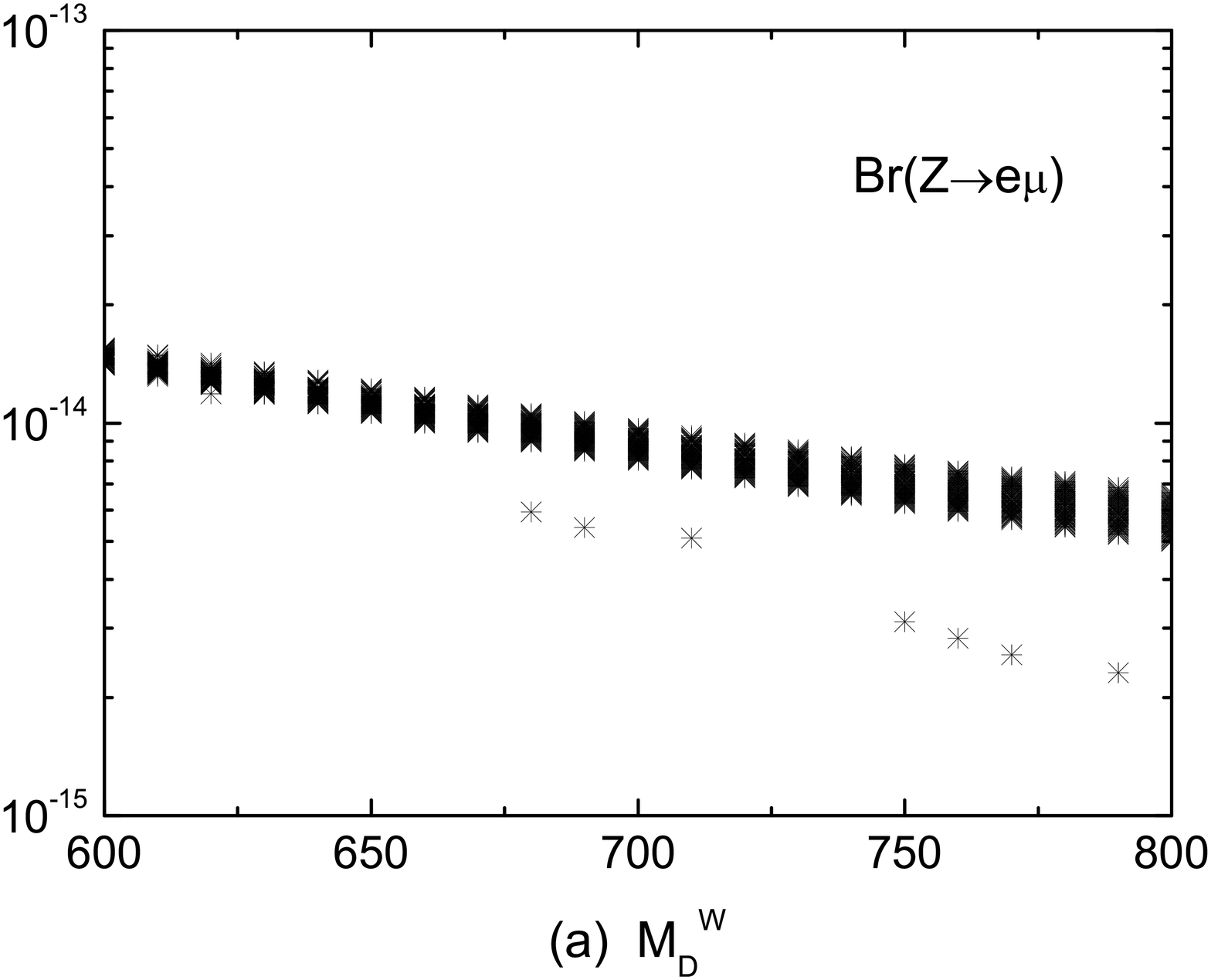}%
\includegraphics[width=0.5\columnwidth]{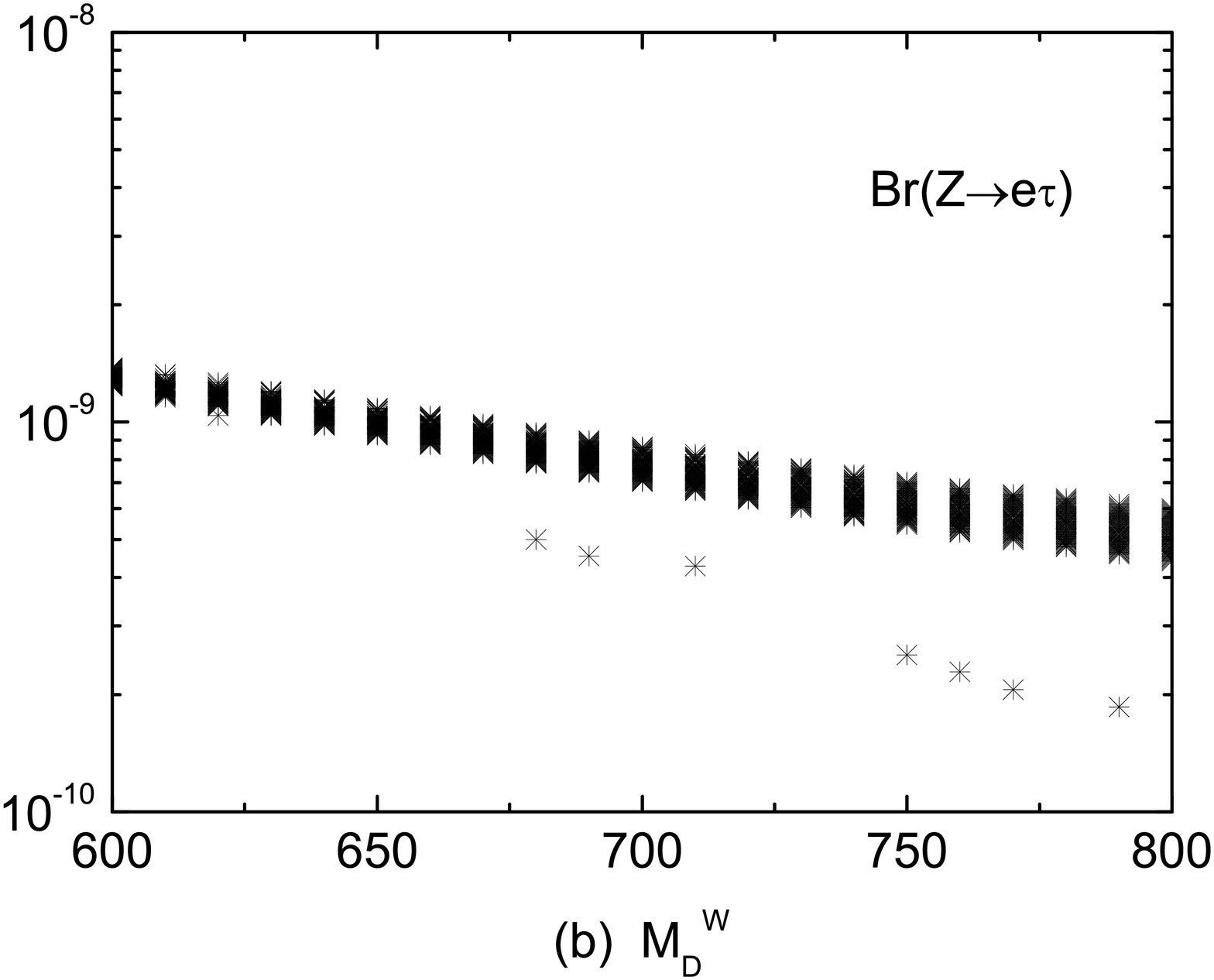}
\includegraphics[width=0.5\columnwidth]{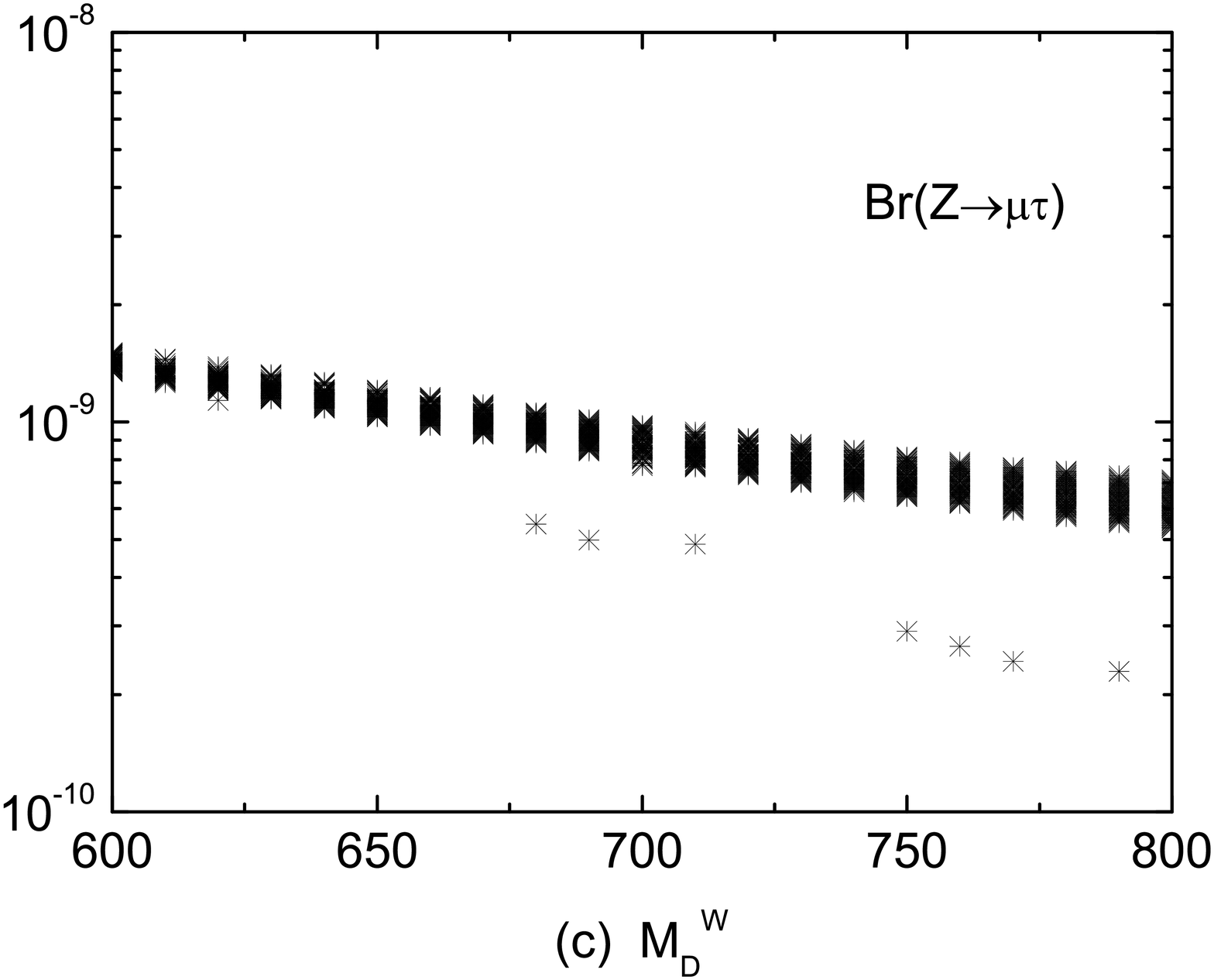}
\end{minipage}
\caption[]{(a)Br($Z\rightarrow e\mu$) versus $M_D^W$; (b)Br($Z\rightarrow e\tau$) versus $M_D^W$; (c)Br($Z\rightarrow \mu\tau$) versus $M_D^W$. All points are satisfied with the experimental bound in TABLE.\ref{lfvl}.}
\label{figs2}
\end{figure}

Recently, the ATLAS collaboration has released a search for chargino-neutralino production
in two and three lepton final states employing RJR techniques that target specific
event topologies \cite{AT1806}, which state that charginos and neutralinos must be heavier than 600 GeV at 95\% CL.
To be compatible with the experimental limit, the parameters $M_D^W$, $M_D^B$, $\mu_u$ and $\mu_d$, which dominant the masses of charginos and neutralinos, should be enlarged. The selection of BMP1 and BMP3 is shown in figure 8.2 in reference \cite{PhD}, where the parameters are set to BMP1 and BMP3 for the top and bottom row respectively and the benchmark point is marked by a star in each plot. It is shown that the valid region is $\mu_u(\mu_d)<$ 500 GeV for BMP1, and this leads to at least two sparticle mass are lighter than 600 GeV. For BMP3, the valid region for $\mu_u(\mu_d)$
can enlarged above 600 GeV together with $M_D^W$ and $M_D^B$. Then, to quantitatively study the LFV decays of Z boson, we perform a scan under BMP3 over the parameters $M_D^W$ and $M_D^B$, with $\mu_u(\mu_d)=$ 600 GeV. The ranges of variation over the MRSSM parameters are displayed in TABLE.\ref{scan}, all mass parameters are in GeV. Moreover, the three flavor violating parameters $\delta^{12}$=$10^{-3}$, $\delta^{13}$ =0.3 and $\delta^{23}$ =0.3 are assumed.

Over a general scan of thousands of points in parameter spaces according to TABLE.\ref{scan}, we display the predictions on Br($Z\rightarrow e\mu$), Br($Z\rightarrow e\tau$) and Br($Z\rightarrow \mu\tau$) as a function of $M_D^W$ in FIG.\ref{figs2}(a), (b) and (c) respectively. It shows that the predictions on Br($Z\rightarrow e\mu$), Br($Z\rightarrow e\tau$) and Br($Z\rightarrow \mu\tau$) decrease as $M_D^W$ varies from 600 GeV to 800 GeV. For Br($Z\rightarrow e\mu$), the range of the prediction is narrowed above $\mathcal{O}(10^{-14})$ at $M_D^W$=600 GeV and narrowed to $\mathcal{O}(10^{-15}-10^{-14})$ at $M_D^W$=800 GeV. For Br($Z\rightarrow e\tau$) and Br($Z\rightarrow \mu\tau$), the range of the prediction is narrowed above $\mathcal{O}(10^{-9})$ at $M_D^W$=600 GeV and  $\mathcal{O}(10^{-10}-10^{-9})$ at $M_D^W$=800 GeV.

\section{Conclusions\label{sec4}}
In this work, taking account of the constraints on the parameter space from radiative charged lepton decays Br($l_2\rightarrow l_1\gamma$), we analyze the LFV decays of $Z\rightarrow l_1 l_2$ as a function of the flavor violating off-diagonal entries $\delta^{12}$, $\delta^{13}$ and $\delta^{23}$ in the framework with R-symmetric Supersymmetric Standard Model. A summary on the upper limits on flavor violating parameters in MRSSM is given in TABLE.\ref{result1}, where the results in MSSM are also included for supplement. It displays the upper limits in two different models are very close to each other.
\begin{table}[h]
\caption{Upper limits on off-diagonal entries $\delta^{12}$, $\delta^{13}$ and $\delta^{23}$. }
\begin{tabular}{@{}cccc@{}} \toprule
Parameters&MRSSM&MSSM($\Delta^{IJ}_{LL}$/$\Delta^{IJ}_{RR}$\cite{Rosiek1})\\
\colrule
$\delta^{12}$&$1.0\times 10^{-3}$&$8.4\times 10^{-4}$/$5.0\times 10^{-3}$\\
$\delta^{13}$&$3.0\times 10^{-1}$&$5.3\times 10^{-1}$/$\mathcal{O}(1)$\\
$\delta^{23}$&$3.0\times 10^{-1}$&$4.6\times 10^{-1}$/$\mathcal{O}(1)$\\
\botrule
\end{tabular}
\label{result1}
\end{table}
\begin{table}[h]
\caption{Upper predictions on LFV decays of $Z$ boson in MRSSM. }
\begin{tabular}{@{}cccc@{}} \toprule
Decay&Prediction&Future sensitivity\cite{Z1,Z2,Z3,Z4}\\
\colrule
Br($Z\rightarrow e\mu$)&$\mathcal{O}(10^{-14})$&$2.0\times 10^{-9}$\\
Br($Z\rightarrow e\tau$)&$\mathcal{O}(10^{-9})$&$(1.3-6.5)\times 10^{-8}$\\
Br($Z\rightarrow \mu\tau$)&$\mathcal{O}(10^{-9})$&$(0.44-2.2)\times 10^{-8}$\\
\botrule
\end{tabular}
\label{result2}
\end{table}

The LFV decays of $Z\rightarrow l_i l_j$ depend strongly on the three flavor violating parameters $\delta^{ij}$ in soft breaking terms $m_l$ and $m_r$, i.e., if set $\delta_{ij}$=0, then the branching ratios of $Z\rightarrow l_i l_j$ equal zero. Taking account of constraints from $l_2\rightarrow l_1\gamma$ we summarize the theoretical predictions of $Z\rightarrow l_1 l_2$ in MRSSM in TABLE.\ref{result2}, and the value is taken from FIG.\ref{figs2} with $M_D^W$=600 GeV. The upper prediction on Br($Z\rightarrow e\mu$) is seven orders of magnitude below current experimental bound and we may make more efforts to observe it in future experiment. The upper prediction on Br($Z\rightarrow e\tau$) and Br($Z\rightarrow \mu\tau$) are at the same order and close to the current experimental bound. Thus, the LFV decays $Z\rightarrow e\tau$ and $Z\rightarrow \mu\tau$ may be more promising to be observed in future experiment.

\begin{acknowledgments}
\indent\indent
The work has been supported by the National Natural Science Foundation of China (NNSFC) with Grants No.11747064, No.11805140, the Scientific and Technological Innovation Programs of Higher Education Institutions in Shanxi (Grant No. 2017113), the Foundation of Department of Education of Liaoning province with Grant No. 2016TSPY10 and the Youth Foundation of the University of Science and Technology Liaoning with Grant No. 2016QN11.

\end{acknowledgments}

\end{document}